\documentclass[aps,twocolumn,prb,showpacs]{revtex4}
\usepackage[dvips]{color}
\usepackage{graphicx}
\usepackage{epsfig}
\usepackage{dcolumn}
\usepackage{color}
\usepackage{amssymb}
\usepackage[english]{babel}
\usepackage{mathrsfs}
\usepackage{bm}
\usepackage{amsmath}
\usepackage{lscape}
\newcommand{\mat}[4]{\left( \begin{array}{cc}
#1 & #2 \\
#3 & #4
\end{array}\right)}
\newcommand{\vect}[2]{\left(\begin{array}{c}#1 \\ #2\end{array}\right)}

\begin{document}

\title{Photon correlations in a two-site non-linear cavity system \\
under coherent drive and dissipation}

\author{Sara Ferretti and Lucio Claudio Andreani}
\affiliation{Dipartimento di Fisica ``Alessandro Volta'',
Universit\`a di Pavia,  via Bassi 6, I-27100 Pavia, Italy}

\author{Hakan E. T\"ureci}
\affiliation{Institute for Quantum Electronics, ETH Zurich, 8093 Zurich, Switzerland}

\author{Dario Gerace}
\affiliation{Dipartimento di Fisica ``Alessandro  Volta'', and UdR CNISM,
Universit\`a di Pavia,  via Bassi 6, I-27100 Pavia, Italy}


\date{\today}

\begin{abstract}
We calculate the normalized second-order correlation function for a
system of two tunnel-coupled photonic resonators, each one
exhibiting a single-photon nonlinearity of the Kerr type. We employ
a full quantum formulation: the master equation for the model, which
takes into account both a coherent continuous drive and radiative as
well as non-radiative dissipation channels, is solved analytically
in steady state through a perturbative approach, and the results are
compared to exact numerical simulations. The degree of second-order
coherence displays values between 0 and 1, and divides the diagram
identified by the two energy scales of the system - the tunneling
and the nonlinear Kerr interaction -  into two distinct regions
separated by a crossover. When the tunneling term dominates over the
nonlinear one, the system state is delocalized over both cavities
and the emitted light is coherent. In the opposite limit, photon
blockade sets in and the system shows an insulator-like state with
photons locked on each cavity, identified by antibunching of emitted
light.
\end{abstract}

\maketitle

\section{Introduction}\label{intro}

Recent advances in cavity quantum electrodynamics (CQED) have led to
the demonstration of a number of striking phenomena related to the
fundamental properties of light-matter coupling, e.g. when single or
few quantum emitters interact with the mode of an electromagnetic
resonator.\cite{raimond01rev,mabuchi02rev,haroche_book} The
experimental realization of the Jaynes-Cummings (JC) model,\cite{jc}
which predicts a strong light-matter coupling regime when the Rabi
frequency between the oscillators exceeds their respective loss
rates,\cite{carmicheal,andreani99prb} has been achieved in both
atomic\cite{raimond01rev} and solid-state\cite{sc_papers,wallraff04}
CQED with single two-level emitters in high-Q resonators. In the
strong coupling regime, the CQED system is intrinsically anharmonic
at the level of single quanta, which derives from the underlying
anharmonic nature of the emitter's
eigenstates.\cite{turchette,schuster08np,kevin07nat,kartik07nat,bishop09np,delvalle09prb}

Following these early works, the ultimate limit of nonlinear optics,
i.e. the ability to control the nonlinear response of a system by
the injection of single photons through the so called \textit{photon
blockade} effect,\cite{imamoglu99} has been recently reached.
Inhibition of the resonant transmission of a single photon because
of the presence of another one within the cavity has been
experimentally demonstrated with both single atoms in optical
cavities\cite{birnbaum05nat} and semiconductor quantum dots strongly
coupled to photonic nanocavities.\cite{faraon08nphys} In all these
experiments, the statistical properties of the resonant light beam
transmitted through the nonlinear system gives precise information
on the nature of the effective photon-photon interaction within the
cavity. Photon blockade has been shown to be strictly characterized
by conversion of a classical, coherent field at the input into a
nonclassical, antibunched photon stream at its
output.\cite{imamoglu99} {Similar effects have been also predicted
for other types of coherent
fields.\cite{carusotto01pra,rebic04pra,carusot09preprint} More
recently, there has been an intense effort towards the exploitation
of nonlinearities arising from Coulomb interaction in confined
electron and photon systems.\cite{deveaud06apl} In such a case,
mixed light-matter states (\textit{polaritons}) arise from the
strong coupling regime of quantum well excitons and microcavity
photons, which are three-dimensionally confined thanks to the
progress in lithographic techniques. The polariton quantum blockade
has been predicted for such highly nonlinear light-matter
states,\cite{ciuti06prb} and first evidences of nonlinear behavior
have been reported experimentally.\cite{deveaud08apl} These systems
are likely to provide a further playground for single-photon
nonlinear optics in the near future.}

Motivated by the great level of control achieved in CQED experiments
with single cavities, recent theoretical work has explored
multi-cavity nonlinear systems. Initial work was mainly aimed at
studying the superfluid-insulator quantum phase
transition\cite{fisher89,jaksch98,greiner02} of the effective
Bose-Hubbard
hamiltonian,\cite{hartmann08,hartmann06,na08prb,aichorn08prl,kay08,pippan09pra}
or the JC-Hubbard
model\cite{greentree06,rossini07,makin,grochol09pra,koch09pra,sebastian09prl,sebastian10prl}
for arrays of CQED systems under quasi-equilibrium conditions.
Subsequent work dealing with coupled non-linear cavity systems has
addressed the dynamics in a two-site JC
model,\cite{ogden08pra,schmidtst10} soliton
physics,\cite{paternostro} a proposal for observing fractional
quantum Hall states,\cite{cho08} the possible realization of a
Tonks-Girardeau gas in different one-dimensional
geometries,\cite{chang08np,carusotto09} the study of effective spin
models\cite{angelakis07,hartmann07} and of entanglement
generation,\cite{cho08pra} {the use of coupled cavity systems as
efficient single-photon sources even in the presence of weak photon
nonlinearities,\cite{savona10prl} and the signatures of
superfluid-insulator quantum phase transition for an infinite CQED
array under pulsed coherent driving.\cite{tomadin10pra} }

In a recent work,\cite{gerace_josephson} a proposal has been made to
observe signatures of strong photon correlations in a system of
three coupled nonlinear cavities, with the central one displaying
single-photon nonlinearity, through the measurement of its degree of
second order coherence. Besides being a readily realizable system
with state-of-the art technology with both atomic and solid state
CQED,\cite{wallraff04,kevin07nat,kapon08oe,vignolini09apl} this
system is a possible quantum photonic device in which information
encoded in classical field states can be controlled by the presence
or absence of single photon quanta (a single-photon transistor).

\begin{figure}[t]
\begin{center}
\includegraphics[width=0.5\textwidth]{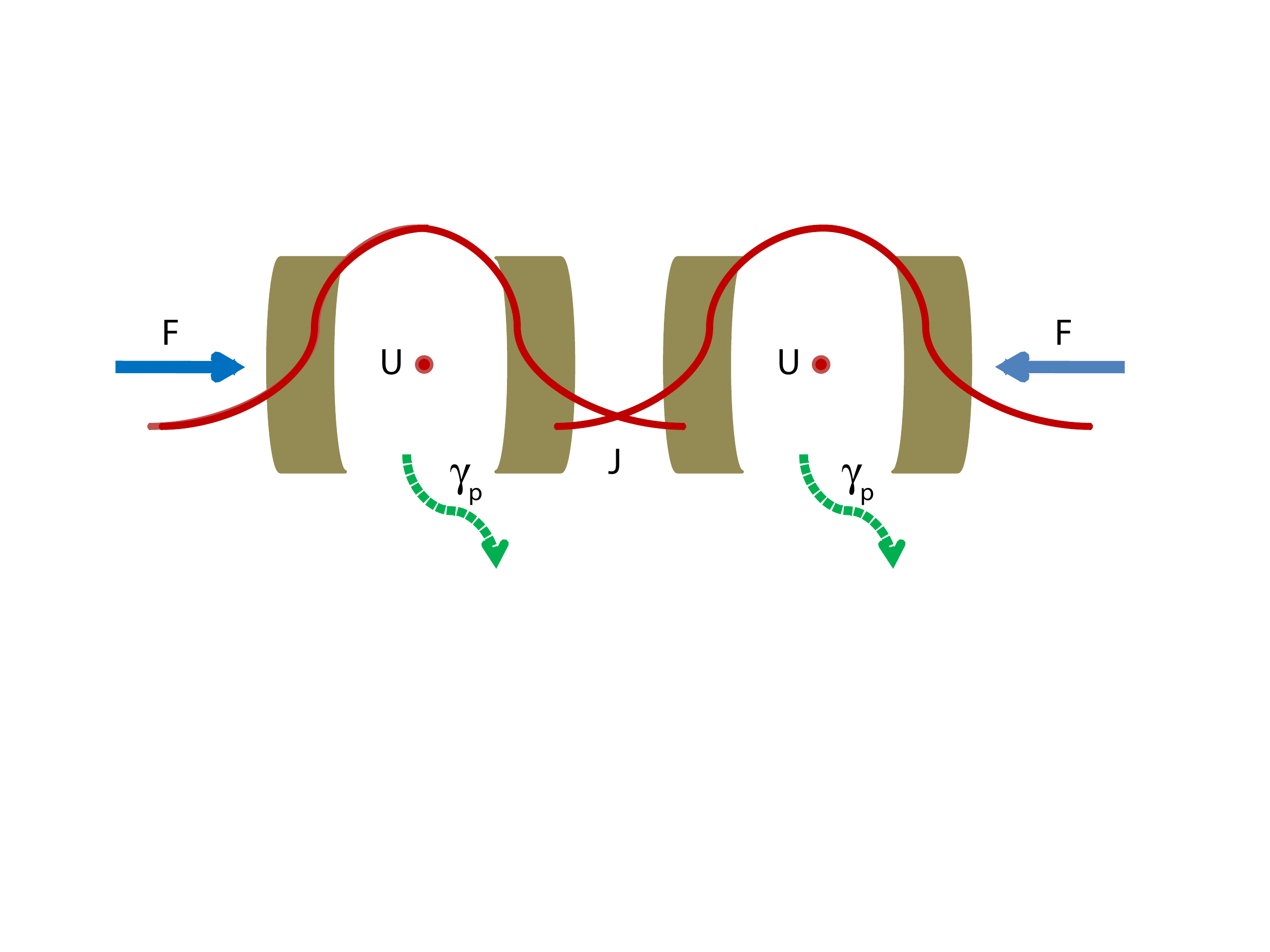}
\vspace{-2.0cm} \caption{(Color online) The system of two coupled
nonlinear cavities. The relevant parameters of the model are
indicated, namely the coherent pumping ($F$) and dissipation
($\gamma_{p}$) rates, respectively. The cavities are supposed to be
nonlinear, with an interaction energy $U$, and they are
tunnel-coupled by evanescent overlap of their cavity modes with a
coupling constant $J$. We assume symmetric cavity parameters in this
work. } \label{fig1}
\end{center}
\end{figure}

In order to extend and generalize the latter work, here we present a
systematic theoretical analysis of a model of two coupled cavities,
\textit{both} of them assumed to be nonlinear at the single photon
level. The  model takes into account coherent driving as well as
global dissipation channels within a master equation treatment (see
a sketch of the system, Fig.~\ref{fig1}). The dynamical equilibrium
reached by the system is due to the balance between pumped and
dissipated photons and it is analyzed in steady state. Following
previous work,\cite{gerace_josephson} we mainly concentrate on
calculating the second-order correlation function for the light
emitted from each cavity, both analytically and numerically.
Interplay of coherent tunneling and on-site interactions is clearly
identified in the crossover from Poissonian to sub-Poissonian light
statistics of the emitted light.

The paper is organized as follows:  we first introduce the model and
the master equation in Sec.~II. In Sec.~III we provide a description
of the analytical solution for the master equation that can capture
the  second-order correlation function for this model, and compare
it to a full numerical solution.

% FIG. 1
%\vspace{-1.5cm}

\section{ Theoretical framework }\label{model}

{We will investigate here the photon correlations in the two-site
\textit{Kerr-Hubbard model} (Fig.~\ref{fig1}) given by ($\hbar=1$)
\begin{eqnarray}\label{H2C_Kerr}
\hat{H}_{\mathrm{KH}} &=& \sum_{i=1,2} [ \omega_{i} \hat{p}^{\dag}_i
\hat{p}_i +U_i \hat p_{i}^{\dag}\hat p_{i}^{\dag}\hat p_{i}\hat
p_{i} + F_i e^{-\mathrm{i}\omega_{\mathrm{L}}t}  \hat{p}_{i}^{\dag}
+ F_i^{\ast} e^{\mathrm{i}\omega_{\mathrm{L}}t} \hat{p}_{i} ]
\nonumber \\
&+& J(\hat{p}_{1}^{\dag}\hat{p}_{2}+\hat p_{2}^{\dag}\hat{p}_{1}) \,
,
\end{eqnarray}
where $\hat{p}_i$ ($\hat{p}_i^{\dag}$) destroy (create) generic
bosonic excitations in each of the two cavities at their fundamental
frequencies $\omega_i$, $U_i$ is the nonlinear Kerr-type interaction
in each cavity, $J$ is the inter-cavity tunneling rate, and $F_i
e^{-\mathrm{i}\omega_{\mathrm{L}}t}$ is the coherent driving
amplitude at laser frequency $\omega_{\mathrm{L}}$.

There are several ways to realize the model in Eq.~(\ref{H2C_Kerr}),
with either atomic or solid-state cavity QED technology. All these
realizations principally rely on the formation of well-defined
quasi-particles, \textit{polaritons}, of mixed light-matter nature,
for the availability of the required effective quasi-particle
interactions $U_{i}$. One possible way is, e.g., to start from two
atoms or quantum dots strongly coupled to their respective cavity
modes, with the two cavities in optical contact with each other.
While the single-photon nonlinearity derives from the light-matter
coupling in the framework of a JC model, the tunnel coupling is due
to photon tunneling and the coherent pump acts directly on the
photonic degrees of freedom. In the absence of losses or spontaneous
emission decay, such a system would be described by a two-site
JC-Hubbard Hamiltonian, extensively discussed in recent
literature.\cite{greentree06,koch09pra,sebastian09prl,sebastian10prl,ogden08pra,schmidtst10}
In the weak pumping limit and the dispersive regime, the JC
nonlinearity can be reduced to an effective Kerr-type nonlinearity
between polaritons.\cite{note_blais}

A more straightforward and conceptually simple way of obtaining a
Kerr-type nonlinearity that is effective at the single
photon/polariton level is to consider solid-state systems in which
the Coulomb interaction is strong enough. In particular, we refer
here to excitons in quantum wells coupled to a single photonic mode
of a micro-resonator where excitons interact via their dipole
field.\cite{ciuti06prb} In such a case, the Hamiltonian for the
two-site system in the rotating wave and electric dipole
approximations is given by $\hat{H}=\hat{H}_0+\hat{H}_1$
with\cite{savona10prl}
\begin{eqnarray}\label{ham:polaritons}
\hat{H}_0 &=& \sum_{i=1,2}  [ \omega_{\mathrm{cav},i}
\hat{a}^{\dag}_i \hat{a}_i + \omega_{\mathrm{x},i} \hat{X}^{\dag}_i
\hat{X}_i + \Omega_{i} (\hat{a}^{\dag}_i \hat{X}_i + \hat{a}_i
\hat{X}^{\dag}_i) ] \\
\hat{H}_1 &=&  \sum_{i=1,2} [ V_i \hat{X}^{\dag}_i \hat{X}^{\dag}_i
\hat{X}_i \hat{X}_i + E_i (t)e^{-\mathrm{i} \omega_{\mathrm{L}}t}
\hat{a}^{\dag}_i + E^{\ast}_i (t)e^{\mathrm{i} \omega_{\mathrm{L}}t}
\hat{a}_i ]  \nonumber \\
&+& j (\hat{a}^{\dag}_1 \hat{a}_2 + \hat{a}^{\dag}_2 \hat{a}_1) \, .
\end{eqnarray}
Here, $ \hat{a}^{\dag}_i$ ($ \hat{a}_i$) creates (destroys) a photon
in cavity $i$ at frequency $\omega_{\mathrm{cav},i}$, while the
operators $\hat{X}_i$ ($\hat{X}^{\dag}_i$) describe excitonic
quasi-particles with energy $\omega_{\mathrm{x},i}$. We assume an
interaction energy $V_i$ between excitons deriving from a
contact-type Coulomb interaction, and the exciton-photon interaction
strength is given by the Rabi frequency $\Omega_{i}$. Cavity photons
are coherently pumped into each cavity with amplitudes
$E_{i}e^{-\mathrm{i}\omega_{\mathrm{L}}t}$, and $j$ is the tunneling
amplitude for photons between neighboring cavities.

With respect to the latter model, polaritonic excitations can be
defined as linear combination of excitons and cavity photons as
\begin{equation}\label{eq:transform}
\vect{\hat{P}_{-,i}}{\hat{P}_{+,i}} =
\mat{u_{i}}{-v_{i}}{v_{i}}{u_{i}} \vect{\hat{X}_i}{\hat{a}_i}
\end{equation}
where the coefficients are\cite{ciuti03sst}
\begin{equation}
u_{i} = \frac{1} {\sqrt{1+\left( \frac{\Omega_{i}}{\omega_{-,i} -
\omega_{\mathrm{cav},i}} \right)^2 }} \,\, ; \,\, v_{i} =
\frac{1}{\sqrt{1+\left( \frac{\omega_{-,i} -
\omega_{\mathrm{cav},i}} {\Omega_{i}} \right)^2 }} \, .
\end{equation}
This transformation then diagonalizes $\hat{H}_0$
 \begin{equation}
 \hat{H}_{0} = \sum_{\sigma=\pm} \sum_{i=1,2} \omega_{\sigma,i}
(\hat{P}_{\sigma,i})^\dagger
 \hat{P}_{\sigma,i}
 \end{equation}
with the lower and upper polariton energies respectively given by
$\omega_{\pm,i}=(\omega_{\mathrm{cav},i}+\omega_{\mathrm{x},i})/2
\pm \sqrt{\Omega_{i}^2 + (\Delta_{i}/2)^2}$, where
$\Delta_{i}=\omega_{\mathrm{cav},i}-\omega_{\mathrm{x},i}$. Assuming
quasi-resonant pumping of the lower polariton level
($\omega_{\mathrm{L}} \sim \omega_{-}$), $\Delta_{i}<0$ and
neglecting non-resonant contributions, the resulting effective
Hamiltonian $\hat{H}$ can be written in the form of
Eq.~(\ref{H2C_Kerr}) with $\hat{p}_{i} = \hat{P}_{-,i}$, $\omega_{i}
=  \omega_{-,i}$, $F_{i} = v_{i} E_{i}$, $J = v_{1}v_{2}\, j$ and
$U_{i} = u_{i}^{4}V_{i}$. Throughout this work we will be assuming
identical sites for simplicity and drop the site indices on the
parameters of the system.

Losses resulting, e.g., from  spontaneous emission decay of the
excitons or cavity photon leakage can be taken into account within
the quantum Master equation in the Born-Markov approximation for the
density matrix of the quasi-particles in the system, which is
expressed in the usual Lindblad form
\begin{equation}\label{master:eq}
    \frac{\partial}{\partial t}\rho= \mathrm{i}[\rho,\tilde{H}_{\mathrm{KH}}]+{\mathcal{L}}(\rho)\,  ,
\end{equation}
where $\tilde{H}_{\mathrm{KH}}$ is the Kerr-Hubbard Hamiltonian
written in the rotating frame with respect to the pump frequency
resonant with the lower polariton mode ($\omega_{-} -
\omega_{\mathrm{L}} = 0$)
\begin{eqnarray}
\tilde{H}_{\mathrm{KH}} &=& \hat{\mathrm{R}}(t)
\hat{H}_{\mathrm{KH}} \hat{\mathrm{R}}^{\dag}(t) = \sum_{i=1,2} [ U
\hat p_{i}^{\dag}\hat p_{i}^{\dag}\hat p_{i}\hat p_{i} +
   F\hat{p}_{i}^{\dag} + F^{\ast} \hat{p}_{i} ] \nonumber \\
&+& J(\hat{p}_{1}^{\dag}\hat{p}_{2}+\hat p_{2}^{\dag}\hat{p}_{1}) \,
, \label{H2C}
\end{eqnarray}
with $\hat{\mathrm{R}}(t)=\exp\{\mathrm{i}\omega_{\mathrm{L}} (\hat
p_{1}^{\dag}\hat p_{1} + \hat p_{2}^{\dag}\hat p_{2} )t\}$. The
Liouvillian can be expressed in the usual Lindblad form
\cite{carmichael_book}
\begin{equation}\label{L2C}
    {\mathcal{L}}=
    \frac{\gamma_{p}}{2} \sum_{i=1,2} [2\hat{p}_{i}\rho\hat{p}_{i}^{\dag}-
    \hat{p}_{i}^{\dag}\hat p_{i}\rho- \rho\hat{p}_{i}^{\dag}\hat p_{i}] \,
    ,
\end{equation}
where $\gamma_p$ is the polariton dissipation rate in each
cavity.\cite{gammap} We will discuss possible realistic
implementation of this model and the required tolerances in
Sec.~\ref{physre}. }

% FIG 2
\begin{figure}[t]
  \begin{center}
   \includegraphics[width=0.5\textwidth]{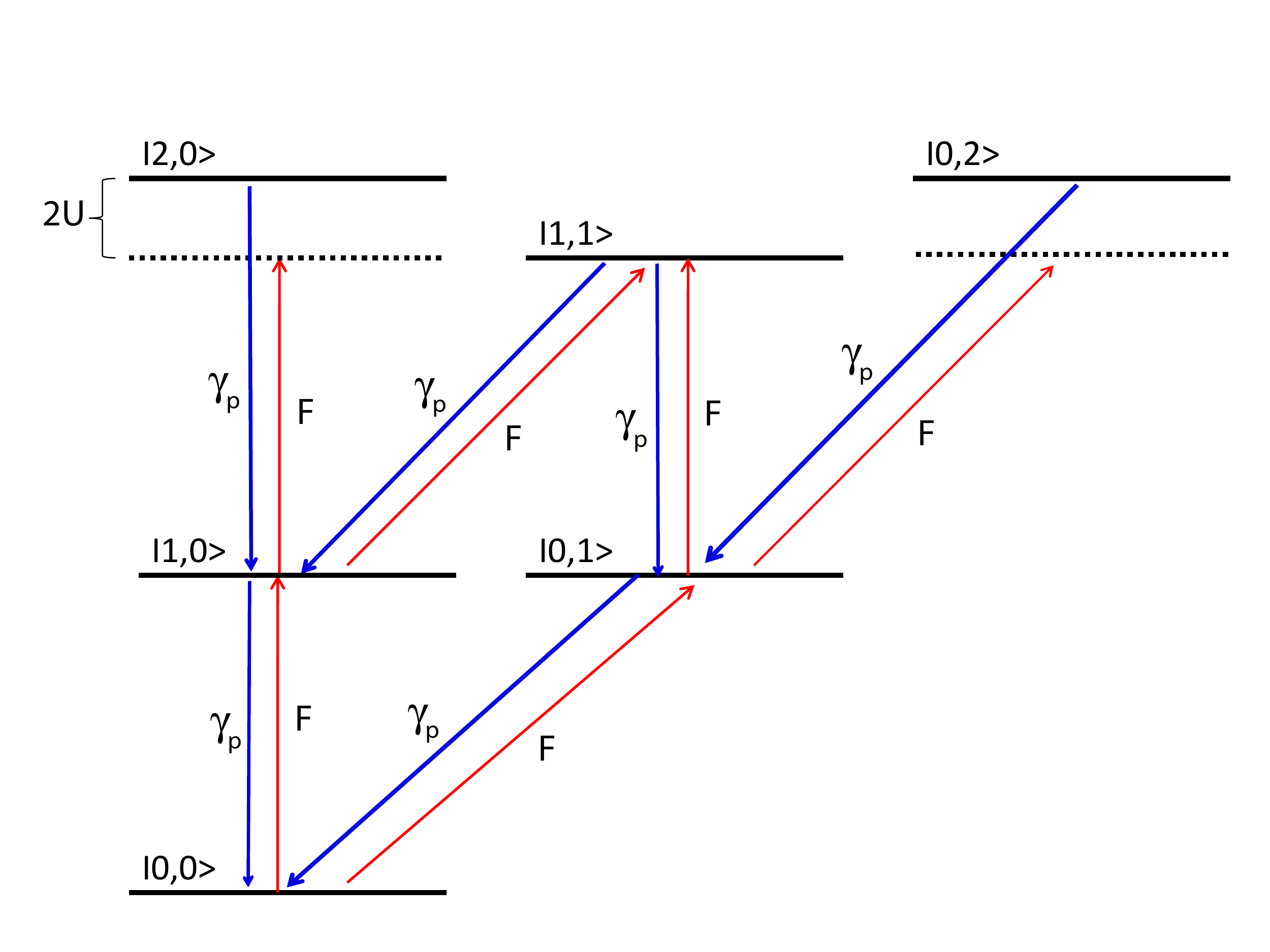}
     \caption{(Color online)
     Schematic energy level diagram and rates for the coupled cavity system.
     We use the following shorthand notation for the eigenstates:
     $|0,0\rangle\rightarrow|1\rangle$,
     $|1,0\rangle\rightarrow|2\rangle$,
     $|0,1\rangle\rightarrow|3\rangle$,
     $|2,0\rangle\rightarrow|4\rangle$,
     $|1,1\rangle\rightarrow|5\rangle$,
     $|0,2\rangle\rightarrow|6\rangle$.}
      \label{fig2}
  \end{center}
\end{figure}

\section{Results}\label{results}

The aim of the present work is to assess the second-order
correlation function as a quantitative probe of the interplay
between tunneling and interactions in a two-site CQED system
described by Eq.~(\ref{master:eq}). To this end, we calculate the
steady state normalized degree of second-order
coherence,\cite{loudon_book} i.e. $g^{(2)}_{ss}(\tau)$, for the
light emitted from the system, which is defined as {
\begin{equation}\label{eq:g2ss}
g^{(2)}_{ss}(\tau)=g^{(2)}(t\rightarrow \infty,\tau)=
\frac{\langle\hat{p}^{\dag}(t)\hat{p}^{\dag}(t+\tau)\hat{p}(t+\tau)\hat{p}(t)\rangle}
{\langle\hat{p}^{\dag}(t)\hat{p}(t)\rangle^2}  \; .
\end{equation}
In the following, we will only be concerned with the zero-time delay
correlation function in steady state, $g^{(2)}_{ss}(0) =
\langle\hat{p}^{\dag2}\hat{p}^{2}\rangle/
\langle\hat{p}^{\dag}\hat{p}\rangle^2$. This correlation function of
the polaritons can be straightforwardly related to the correlation
function of cavity photons via Eqs.~(\ref{eq:transform}), which is
the quantity ultimately measured in a typical
experiment.\cite{gerace_josephson}}

%\subsection{Analytic solution}\label{subsec:analytical}

In the weak pumping limit, an analytic solution to the steady state
master equation for our model can be found in the following way. We
write Eqs.~(\ref{master:eq}) and (\ref{L2C}) in the Fock basis $\{
|n_{1},n_{2}\rangle \}$, where $n_1$ and $n_2$ indicate polariton
occupations in cavities 1 and 2, respectively.  We consider the low
energy excitations of the Hamiltonian (\ref{H2C}),
$N_{tot}=n_{1}+n_{2} \leq 2$. The corresponding energy level diagram
and rates are schematically shown in Fig.~\ref{fig2}. The resulting
equations of motion for the 36 elements of the density matrix can be
solved using perturbation theory and a recursive procedure in
$F/\gamma_{p}$, as described in App.~A.

{The steady-state second-order correlation function
$g^{(2)}_{ss}(0)$ for the $i$th cavity can be calculated as
\begin{equation}\label{eqg2fockrep}
   g_{i}^{(2)}(0)=\frac{\mathrm{Tr}\{ \hat{p}_{i}^{\dag}\hat{p}_{i}^{\dag}\hat{p}_{i}\hat{p}_{i}
\rho_{\mathrm{ss}}\} } {[\mathrm{Tr}\{
\hat{p}_{i}^{\dag}\hat{p}_{i}\rho_{\mathrm{ss}}\}]^2}=\frac{\sum_{mm'} \rho^{\mathrm{ss}}_{m,m'} \langle
   m|\hat{p}_{i}^{\dag}\hat{p}_{i}^{\dag}\hat{p}_{i}\hat{p}_{i}|m'\rangle}
   {\left[ \sum_{m,m'} \rho^{\mathrm{ss}}_{m,m'} \langle m|\hat{p}_{i}^{\dag}\hat{p}_{i}|m'\rangle \right]^{2} }\;,
\end{equation}
where $|m\rangle \equiv |n_{1}n_{2}\rangle$ is a collective notation
for the eigenstates of the coupled cavity system and
$\rho^{\mathrm{ss}}_{m,m'}$ is the steady-state density matrix
calculated from Eqs.~(\ref{master:eq}). We will henceforth drop the
subscript ``ss''. } In the weak pumping limit $F/\gamma_{p} \ll 1$
\begin{equation}\label{eq:g20_an}
    g^{(2)}_1(0)=g^{(2)}_2(0)=g^{(2)}(0)\cong\frac{2{\rho}_{4,4}}{{\rho}_{2,2}^{2}}\;.
\end{equation}
The calculations are lengthy but straightforward (see App.~A), and
the analytic expressions for the matrix elements ${\rho}_{4,4}$ and
${\rho}_{2,2}$ are given in Eqs.~(\ref{rho442cav}) and
(\ref{rho222cav}), respectively. Notice that the explicit analytic
expression for  ${\rho}_{4,4}$ is recursively obtained through the
analytic expression for the elements in Eqs. (\ref{rho222cav}),
(\ref{rho14cav}), (\ref{rho15cav}), (\ref{rho24cav}), and
(\ref{rho25cav}). The procedure can be generalized, e.g., to
perturbatively calculate $g^{(2)}(0)$ for generic multi-site CQED
systems, either analytically or numerically through the
implementation of this recursive algorithm. We notice that setting
$J=0$ in the above elements, Eq. (\ref{eq:g20_an}) gives the correct
limit of the single cavity photon blockade
\begin{equation} \label{eq:singlecav}
    g^{(2)}(0)\cong \frac{1}{1+4(U/\gamma_{p})^{2}}\;.
\end{equation}

\begin{figure}[t]
  \begin{center}
    \includegraphics[width=0.5\textwidth]{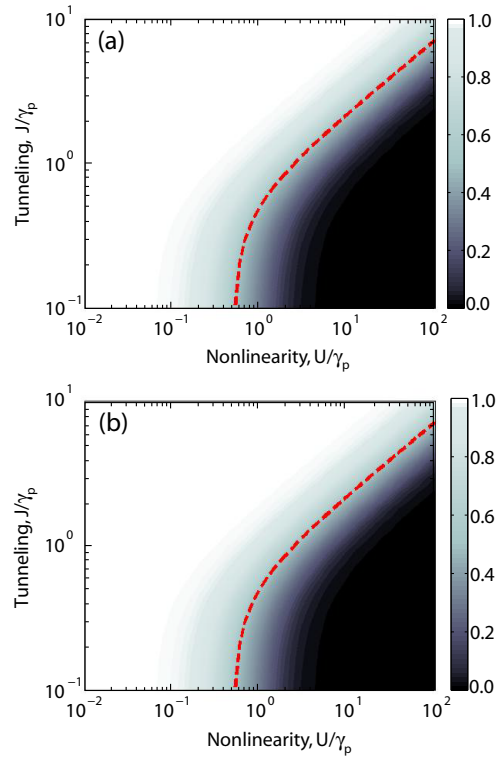}
     \caption{(Color online)
     The second-order correlation function for the two coupled nonlinear cavities,
     $g^{(2)}(0)$, as a function of $U/\gamma_{p}$ and
     $J/\gamma_{p}$ (a) as derived analytically from Eq. (\ref{eq:g20_an}) and
     (b) from numerical solution of Eq.~\ref{master:eq}, calculated for $F/\gamma_{p} = 0.1$.
     {The dashed lines in both (a) and (b) show the boundary curve
     $U_{b}(J)$ defined by the condition $2\rho_{4,4}/ \rho_{2,2}^{2}=0.5$}.}
      \label{fig3}
  \end{center}
\end{figure}

For $J\neq 0$, the resulting behavior of $g^{(2)}(0)$ is shown in
Fig.~\ref{fig3} as a function of dimensionless quantities
$U/\gamma_{p}$ and $J/\gamma_{p}$. The most striking feature of this
plot is the sharp boundary that divides regions where $g^{(2)}(0)
\approx 0$ and $g^{(2)}(0) \approx 1$. {As clearly seen in
Fig.~\ref{fig3}, when the tunneling term dominates over the on-site
interaction energy the emitted light is Poissonian, $g^{(2)}(0)
\approx 1$. This reflects the statistics of the coherent driving
fields imposed on the system due to the dominance of the linear
tunneling terms over the non-linear interaction terms in the
Hamiltonian. Hence the system state is coherently delocalized over
both cavities with symmetric and antisymmetric combinations of the
bare polariton modes. In the opposite limit, the emitted photons are
antibunched, $g^{(2)}(0) \approx 0$, which is a clear indication
that the number of quasi-particles in each cavity can only fluctuate
between zero and one.  The latter is the photon-blockade regime that
would be present for $J=0$ (i.e. individual cavities) for $U \gg
\gamma_{p}$ (See Eq.~(\ref{eq:singlecav})).  Note that for $J \neq
0$ the cross-over takes place for larger values of $U$ as $J$ is
increased, thus showing the interplay of tunneling and interactions
in the steady state. The boundary of the crossover in $U$,
$U_{b}(J)$, can be clearly identified following the $g^{(2)}(0) =
0.5$ contour (see red dashed line in the plot) and is found to
increase approximately linearly with $J$ for $J,U > \gamma_{p}$.  }

% FIG 3

%\subsection{Numerical solution }\label{subsec:numerical}

\begin{figure}[t]
  \begin{center}
     \includegraphics[width=0.5\textwidth]{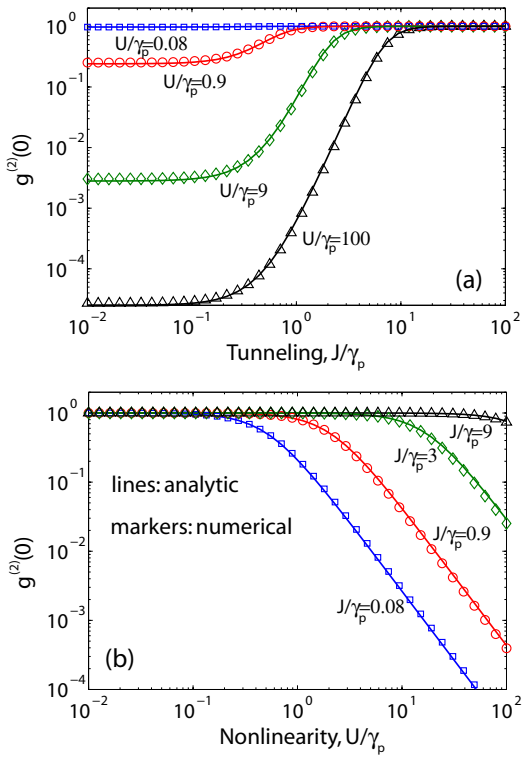}
     \caption{(Color online)
     A direct comparison of numerical and analytic results for $g^{(2)}(0)$:
     (a) fixed $U/ \gamma_{p}$ and varying $J/ \gamma_{p}$, and (b) fixed $J/ \gamma_{p}$
     and varying $U/  \gamma_{p}$. The full lines represent the analytic solutions, while
     numerical ones are shown by symbols.}
      \label{fig4}
  \end{center}
\end{figure}

In order to check the accuracy of the analytical solution presented
above, we have numerically solved
Eqs.~(\ref{master:eq})-(\ref{L2C}). The second order correlation
function is numerically calculated using Eq.~(\ref{eqg2fockrep}), by
using up to 6 photons in the Fock basis to ensure full convergence
(a small $F/ \gamma_{p}$ is assumed to compare with the perturbative
analytic solution). The result is shown in Fig.~\ref{fig3}(b) with a
color scale plot, displaying $g^{(2)}(0)$ as a function of $U/
\gamma_{p}$ and $J/ \gamma_{p}$ to be directly compared to
Fig.~\ref{fig3}(a). To better show the agreement and the accuracy of
the analytical procedure provided in this work, we give in
Fig.~\ref{fig4} several cuts of the color plots of Fig.~\ref{fig3}.
The quantitative behavior  of $g^{(2)}(0)$ is very well reproduced
by the analytic solution over several decades considered in the
parameter space.

% FIG 4

\section{Physical Realization}\label{physre}

{The Kerr-Hubbard Hamiltonian Eq.~(\ref{H2C_Kerr}) can be realized
with a number of systems, including strongly-coupled atom-cavity
systems,\cite{hartmann06} circuit QED systems,\cite{koch09pra} or
quantum dots coupled to semiconductor resonators.\cite{greentree06}

Here, we discuss possible realization with semiconductor-based
optical microcavities. Efficient coupling of optical modes in
photonic resonators in evanescent contact with each other has been
already demonstrated in a number of different systems, including
micro-pillars,\cite{hetterich} micro-disks,\cite{rastelli} and
photonic crystal cavities.\cite{kapon08oe,vignolini09apl} For the
latter, the constant improvement of post-fabrication techniques that
allow to deterministically tune the cavity
modes\cite{kevin06apl,rastelli09} as well as their coupling to
realize the regime of symmetric sites studied here. Note that the
tolerance in the parameters for observing qualitatively the
phenomena studied here in the symmetric limit is given by the loss
rate $\gamma_{p}$. A wide range of tunability has been
experimentally demonstrated in such systems, e.g. for tunnel
coupling rates ($J\sim 0-1$ meV),\cite{vignolini09apl} and Q-factor
engineering allows cavity loss rates to be limited to few
$\mu$eV.\cite{derossi08apl}

Concerning the nonlinearity, as we have seen in Sec.~\ref{model} the
effective Kerr-Hubbard model can be realized either with strongly
coupled two-level systems (qubits) and cavity modes, or through
coupled light-matter states (polaritons) with contact-type
interaction of Coulomb nature. In the first case, sizeable values of
the nonlinearity have been already shown experimentally in
solid-state systems,\cite{kevin07nat,faraon08nphys,bishop09np} and
tunability might be achieved through external control of the
qubit-cavity detuning (see, e.g., Supplementary Information Section
of Ref. \onlinecite{gerace_josephson}). In the second case, for
which photonic crystals represent an ideal platform for a
prospective implementation,\cite{gerace07prb} very large values of
exciton-exciton interactions in confined systems have been predicted
for single resonators,\cite{ciuti06prb} and experimentally measured
recently.\cite{voros09prl}

Finally, detection of emitted radiation and subsequent measurement
of correlation functions would be performed by standard
Hanbury-Brown-Twiss technique, details depending on the geometry of
the system under consideration. For example, in the case of photonic
crystal slab cavity systems emitted light might either be guided
through access waveguides and then detected in an ``in-plane''
geometry, or directly imaged in a camera above the slab as in usual
micro-photoluminescence experiments. Recently, improved far-field
collection from high-Q photonic crystal cavity modes has been
reported,\cite{derossi09prb} which might also be crucial for the
present proposal.}

\section{Conclusions}\label{concl}

The two-site Kerr-Hubbard model represents a paradigmatic system to
investigate the interplay of tunneling and strong correlations in a
CQED system driven out-of-equilibrium. We have analyzed this system
under coherent drive and dissipation and showed that in the
weak-pumping limit a simple perturbative-recursive approach can be
used to analytically compute the system density matrix elements as
well as correlation functions. Such procedure can in principle be
generalized to CQED arrays with more than two sites. The analytical
solution was shown to be highly accurate over several decades of
system parameters by comparing to the numerical solution of the
master equation for the system density matrix.

{Our results show that the zero-delay second-order correlation
function provides an effective probe that can discriminate between
coherent and strongly correlated regimes of a two-site CQED system.
An interesting extension of the present work is the generalization
of the techniques and concepts used here to larger multi-site
systems.}

\begin{acknowledgments}
We thank R. Fazio and S. Schmidt for discussions. This work was
supported from Fondazione Cariplo under grant n. 2007-5259. H.E.T.
acknowledges support from the Swiss NSF under Grant No.
PP00P2-123519/1.

\end{acknowledgments}

\appendix

\begin{widetext}

\section{Density matrix elements}\label{app:elements}

We provide below the analytic solution of
Eqs.~(\ref{master:eq})-(\ref{L2C}) in the small $F/ \gamma_{p}$
limit, where we can truncate the polaritonic Hilbert space to
$N_{tot} = 2$ and assume that the vacuum state to have approximately
unit occupancy, i.e. $\rho_{1,1}\simeq1$. The rest of the density
matrix elements in steady state (see caption to Fig.~\ref{fig2} for
the labelling of the basis states) is given below:
\begin{eqnarray}
    &\rho_{1,1}\simeq1\;,\\
    &\rho_{1,2}=\rho_{2,1}^{*}=\frac{2\texttt{i}}{\gamma_{p}}[J\rho_{1,3}+F(\sqrt{2}\rho_{1,4}+\rho_{1,1}+\rho_{1,5}-\rho_{2,2}-\rho_{3,2})]+2\sqrt{2}\rho_{2,4}+2\rho_{3,5}\;,\\
    &\rho_{1,3}=\rho_{3,1}^{*}=\frac{2\texttt{i}}{\gamma_{p}}[J\rho_{1,2}+F(\rho_{1,5}+\sqrt{2}\rho_{1,6}+\rho_{1,1}-\rho_{2,3}-\rho_{3,3})]+2\rho_{2,5}+2\sqrt{2}\rho_{3,6}\;,\\
    &\rho_{1,4}=\rho_{4,1}^{*}=\frac{\texttt{i}}{\gamma_{p}-2\texttt{i}U}[\sqrt{2}J\rho_{1,5}+F(\sqrt{2}\rho_{1,2}-\rho_{2,4}-\rho_{3,4})]\;,\\
    &\rho_{1,5}=\rho_{5,1}^{*}=\frac{\texttt{i}}{\gamma_{p}}[\sqrt{2}J(\rho_{1,4}+\rho_{1,6})+F(\rho_{1,3}+\rho_{1,2}-\rho_{2,5}-\rho_{3,5})]\;,\\
    &\rho_{1,6}=\rho_{6,1}^{*}=\frac{\texttt{i}}{\gamma_{p}-2\texttt{i}U}[\sqrt{2}J\rho_{1,5}+F(\sqrt{2}\rho_{1,3}-\rho_{2,6}-\rho_{3,6})]\;,\\
    &\rho_{2,2}=\frac{\texttt{i}}{\gamma_{p}}[J(\rho_{2,3}-\rho_{3,2})+F(\sqrt{2}\rho_{2,4}+\rho_{2,1}+\rho_{2,5}-\rho_{1,2}-\sqrt{2}\rho_{4,2}-\rho_{5,2})]+2\rho_{4,4}+\rho_{5,5}\;,\\
    &\rho_{2,3}=\rho_{3,2}^{*}=\frac{\texttt{i}}{\gamma_{p}}[J(\rho_{2,2}-\rho_{3,3})+F(\sqrt{2}\rho_{2,6}+\rho_{2,1}+\rho_{2,5}-\rho_{1,3}-\sqrt{2}\rho_{4,3}-\rho_{5,3})]+2\rho_{4,4}+\rho_{5,5}\;,\\
    &\rho_{3,3}=\frac{\texttt{i}}{\gamma_{p}}[J(\rho_{3,2}-\rho_{2,3})+F(\sqrt{2}\rho_{3,6}+\rho_{3,1}+\rho_{3,5}-\rho_{1,3}-\sqrt{2}\rho_{6,3}-\rho_{5,3})]+2\rho_{4,4}+\rho_{5,5}\;,\\
    &\rho_{2,4}=\rho_{4,2}^{*}=\frac{\texttt{i}}{\frac{3}{2}\gamma_{p}-2\texttt{i}U}[J(\sqrt{2}\rho_{2,5}-\rho_{3,4})+F(\sqrt{2}\rho_{2,2}-\rho_{1,4}-\sqrt{2}\rho_{4,4}-\rho_{5,4})]\;,\\
    &\rho_{2,5}=\rho_{5,2}^{*}=\frac{2}{3}\frac{\texttt{i}}{\gamma_{p}}[J(\sqrt{2}\rho_{2,4}+\sqrt{2}\rho_{2,6}-\rho_{3,5})+F(\rho_{2,3}+\rho_{2,2}-\rho_{1,5}-\sqrt{2}\rho_{4,5}-\rho_{5,5})]\;,\\
    &\rho_{2,6}=\rho_{6,2}^{*}=\frac{\texttt{i}}{\frac{3}{2}\gamma_{p}-2\texttt{i}U}[J(\sqrt{2}\rho_{2,5}-\rho_{3,6})+F(\sqrt{2}\rho_{2,3}-\rho_{1,6}-\sqrt{2}\rho_{4,6}-\rho_{5,6})]\;,\\
    &\rho_{3,4}=\rho_{4,3}^{*}=\frac{\texttt{i}}{\frac{3}{2}\gamma_{p}-2\texttt{i}U}[J(\sqrt{2}\rho_{3,5}-\rho_{2,4})+F(\sqrt{2}\rho_{3,2}-\rho_{5,4}-\rho_{1,4}-\sqrt{2}\rho_{6,4})]\;,\\
    &\rho_{3,5}=\rho_{5,3}^{*}=\frac{2}{3}\frac{\texttt{i}}{\gamma_{p}}[J(\sqrt{2}\rho_{3,4}+\sqrt{2}\rho_{3,6}-\rho_{2,5})+F(\rho_{3,3}+\rho_{3,2}-\rho_{5,5}-\rho_{1,5}-\sqrt{2}\rho_{6,5})]\;,\\
    &\rho_{3,6}=\rho_{6,3}^{*}=\frac{\texttt{i}}{\frac{3}{2}\gamma_{p}-2\texttt{i}U}[J(\sqrt{2}\rho_{3,5}-\rho_{2,6})+F(\sqrt{2}\rho_{3,3}-\rho_{5,6}-\sqrt{2}\rho_{6,6}-\rho_{1,6})]\;,\\
    &\rho_{4,4}=\frac{\texttt{i}}{2\gamma_{p}}[\sqrt{2}J(\rho_{4,5}-\rho_{5,4})+\sqrt{2}F(\rho_{4,2}-\rho_{2,4})]\;,\\
    &\rho_{4,5}=\rho_{5,4}^{*}=\frac{\texttt{i}}{2\gamma_{p}+2\texttt{i}U}[\sqrt{2}J(\rho_{4,4}+\rho_{4,6}-\rho_{5,5})+F(\rho_{4,3}+\rho_{4,2}-\sqrt{2}\rho_{2,5})]\;,\\
    &\rho_{4,6}=\rho_{6,4}^{*}=\frac{\texttt{i}}{2\gamma_{p}}[\sqrt{2}J(\rho_{4,5}-\rho_{5,6})+\sqrt{2}F(\rho_{4,3}-\rho_{2,6})]\;,\\
    &\rho_{5,5}=\frac{\texttt{i}}{2\gamma_{p}}[\sqrt{2}J(\rho_{5,4}+\rho_{5,6}-\rho_{6,5}-\rho_{4,5})+F(\rho_{5,3}+\rho_{5,2}-\rho_{3,5}-\rho_{2,5})]\;,\\
    &\rho_{5,6}=\rho_{6,5}^{*}=\frac{\texttt{i}}{2\gamma_{p}-2\texttt{i}U}[\sqrt{2}J(\rho_{5,5}-\rho_{6,6}-\rho_{4,6})+F(\sqrt{2}\rho_{5,3}-\rho_{3,6}-\rho_{2,6})]\;,\\
    &\rho_{6,6}=\frac{\texttt{i}}{2\gamma_{p}}[\sqrt{2}J(\rho_{6,5}-\rho_{5,6})+\sqrt{2}F(\rho_{6,3}-\rho_{3,6})]\;,
\end{eqnarray}
%\end{widetext}

We simplify the equations above as follows. First we consider the
equations with a first order dependence on $F/\gamma_{p}$, and we neglect
all the higher order terms. We get
\begin{equation}
   \rho_{1,2}=\rho_{2,1}^{*}=\frac{2\texttt{i}}{\gamma_{p}}(J\rho_{1,3}+F\rho_{1,1})\;,
\end{equation}
and
\begin{equation}
   \rho_{1,3}=\rho_{3,1}^{*}=\frac{2\texttt{i}}{\gamma_{p}}(J\rho_{1,2}+F\rho_{1,1})\;.
\end{equation}
This yields
%\begin{widetext}
\begin{equation}
    \rho_{1,2}=\rho_{1,3}=\rho_{2,1}^{*}=\rho_{3,1}^{*}=\frac{\frac{2\texttt{i}}{\gamma_{p}}(1+\frac{2\texttt{i}J}{\gamma_{p}})}{1+(\frac{2J}{\gamma_{p}})^{2}}F\rho_{1,1}\;.
\end{equation}
%\end{widetext}
Next we insert these expressions into the remaining equations to obtain equations at a higher order in $F/\gamma_{p}$. Thus, the elements  of $ \rho$ with a second order dependence on
$F/\gamma_{p}$ give the following set of closed equations
%\begin{widetext}
\begin{eqnarray}
    &\rho_{2,2}=\frac{\texttt{i}}{\gamma_{p}}[J(\rho_{2,3}-\rho_{3,2})+F(\rho_{2,1}-\rho_{1,2})]=\frac{\texttt{i}}{\gamma_{p}}F(\rho_{2,1}-\rho_{1,2})\; \\
    &\rho_{2,3}=\frac{\texttt{i}}{\gamma_{p}}[J(\rho_{2,2}-\rho_{3,3})+F(\rho_{2,1}-\rho_{1,3})]=\frac{\texttt{i}}{\gamma_{p}}F(\rho_{2,1}-\rho_{1,3})\; \\
    &\rho_{3,2}=\frac{\texttt{i}}{\gamma_{p}}[J(\rho_{3,3}-\rho_{2,2})+F(\rho_{3,1}-\rho_{1,2})]=\frac{\texttt{i}}{\gamma_{p}}F(\rho_{3,1}-\rho_{1,2})\; \\
    &\rho_{3,3}=\frac{\texttt{i}}{\gamma_{p}}[J(\rho_{3,2}-\rho_{2,3})+F(\rho_{3,1}-\rho_{1,3})]=\frac{\texttt{i}}{\gamma_{p}}F(\rho_{3,1}-\rho_{1,3})\; ,
\end{eqnarray}
%\end{widetext}
from which we can calculate a solution in terms of $\rho_{1,1}$ as
%\begin{widetext}
\begin{eqnarray}\label{rho222cav}
    &\rho_{2,2}=\rho_{2,3}=\rho_{3,2}=\rho_{3,3}=\frac{4}{1+(\frac{2J}{\gamma_{p}})^{2}}(\frac{F}{\gamma_{p}})^{2}\rho_{1,1}\;.
\end{eqnarray}
%\end{widetext}
Now we consider the equations of order $(F/ \gamma_{p})^{3}$
%\begin{widetext}
\begin{eqnarray}
    &\rho_{1,4}=\rho_{4,1}^{*}=\frac{\sqrt{2}\texttt{i}J}{\gamma_{p}-2\texttt{i}U}\rho_{1,5}+\frac{\sqrt{2}\texttt{i}F}{\gamma_{p}-2\texttt{i}U}\rho_{1,2}\; \\
    &\rho_{1,6}=\rho_{6,1}^{*}=\frac{\sqrt{2}\texttt{i}J}{\gamma_{p}-2\texttt{i}U}\rho_{1,5}+\frac{\sqrt{2}\texttt{i}F}{\gamma_{p}-2\texttt{i}U}\rho_{1,3}\; \\
    &\rho_{1,5}=\rho_{5,1}^{*}=2\sqrt{2}\texttt{i}\frac{J}{\gamma_{p}}\rho_{1,4}+2\texttt{i}\frac{F}{\gamma_{p}}\rho_{1,2} \; .
\end{eqnarray}
%\end{widetext}
Again, we can solve the set of coupled equations reported above
isolating the explicit dependence on $\rho_{1,1}$, which gives the solutions
%\begin{widetext}
\begin{eqnarray}
    &\rho_{1,4}=\rho_{1,6}=\rho_{4,1}^{*}=\rho_{6,1}^{*}=\frac{-2\sqrt{2}(1+\frac{2\texttt{i}J}{\gamma_{p}})^{2}}{[\gamma_{p}(\gamma_{p}-2\texttt{i}U)+4J^{2}][1+(\frac{2J}{\gamma_{p}})^{2}]}F^{2}\rho_{1,1}\;,
    \label{rho14cav}\\
    &\rho_{1,5}=\rho_{5,1}^{*}=-\frac{4\texttt{i}J+2(\gamma_{p}-2\texttt{i}U)}{4J^{2}+\gamma_{p}(\gamma_{p}-2\texttt{i}U)}\frac{\frac{2}{\gamma_{p}}(1+\frac{2\texttt{i}J}{\gamma_{p}})}{1+(\frac{2J}{\gamma_{p}})^{2}}F^{2}
    \rho_{1,1}
    \label{rho15cav}\; .
\end{eqnarray}
%\end{widetext}
We analyse the other set of equations of order $(F/\gamma_{p})^3$
%\begin{widetext}
\begin{eqnarray}
    &\rho_{2,4}=\frac{\texttt{i}}{\frac{3}{2}\gamma_{p}-2\texttt{i}U}[J(\sqrt{2}\rho_{2,5}-\rho_{2,6})+F(\sqrt{2}\rho_{2,2}-\rho_{1,4})]\;,\\
    &\rho_{2,6}=\frac{\texttt{i}}{\frac{3}{2}\gamma_{p}-2\texttt{i}U}[J(\sqrt{2}\rho_{2,5}-\rho_{3,6})+F(\sqrt{2}\rho_{2,3}-\rho_{1,6})]\;,\\
    &\rho_{3,4}=\frac{\texttt{i}}{\frac{3}{2}\gamma_{p}-2\texttt{i}U}[J(\sqrt{2}\rho_{3,5}-\rho_{2,4})+F(\sqrt{2}\rho_{3,2}-\rho_{1,4})]\;,\\
    &\rho_{3,6}=\frac{\texttt{i}}{\frac{3}{2}\gamma_{p}-2\texttt{i}U}[J(\sqrt{2}\rho_{3,5}-\rho_{2,6})+F(\sqrt{2}\rho_{3,3}-\rho_{1,6})]\;,\\
    &\rho_{2,5}=\frac{2}{3}\frac{\texttt{i}}{\gamma_{p}}[J(\sqrt{2}\rho_{2,4}+\sqrt{2}\rho_{2,6}-\rho_{3,5})+F(\rho_{2,3}+\rho_{2,2}-\rho_{1,5})]\;,\\
    &\rho_{3,5}=\frac{2}{3}\frac{\texttt{i}}{\gamma_{p}}[J(\sqrt{2}\rho_{3,4}+\sqrt{2}\rho_{3,6}-\rho_{2,5})+F(\rho_{3,3}+\rho_{3,2}-\rho_{1,5})]\;,
\end{eqnarray}
%\end{widetext}
and we get a solution depending on the elements $\rho_{2,2}$,
$\rho_{1,4}$, $\rho_{1,5}$, which in turn depend on $\rho_{1,1}$ as shown above. The solutions for $\rho_{2,4}$ and $\rho_{2,5}$ read
%\begin{widetext}
\begin{eqnarray}
    &\rho_{2,4}=\rho_{3,6}=\rho_{2,6}=\rho_{3,4}=\rho_{4,2}^{*}=\rho_{6,3}^{*}=\rho_{6,2}^{*}=\rho_{4,3}^{*}=\texttt{i}F\frac{(3\gamma_{p}+2\texttt{i}J)(\sqrt{2}\rho_{2,2}-\rho_{1,4})+2\sqrt{2}\texttt{i}J(2\rho_{2,2}-\rho_{1,5})}{(\frac{3}{2}\gamma_{p}-2\texttt{i}U)(3\gamma_{p}+2\texttt{i}J)+3\texttt{i}\gamma_{p}
    J+6J^{2}}\;, \label{rho24cav} \\
    &\rho_{2,5}=\rho_{3,5}=\rho_{5,2}^{*}=\rho_{5,3}^{*}=\texttt{i}F\frac{4\sqrt{2}\texttt{i}J(\sqrt{2}\rho_{2,2}-\rho_{1,4})+2[(\frac{3}{2}\gamma_{p}-2\texttt{i}U)+\texttt{i}J](2\rho_{2,2}-\rho_{1,5})}{(\frac{3}{2}\gamma_{p}-2\texttt{i}U)(3\gamma_{p}+2\texttt{i}J)+3\texttt{i}\gamma_{p} J+6J^{2}}
    \label{rho25cav}\; .
\end{eqnarray}
%\end{widetext}
We finally  consider the terms depending on $(F/\gamma_{p})^{4}$,
%\begin{widetext}
\begin{eqnarray}
    &\rho_{4,5}=\rho_{5,4}^{*}=\frac{\texttt{i}}{2\gamma_{p}+2\texttt{i}U}[\sqrt{2}J(2\rho_{4,4}-\rho_{5,5})+F(2\rho_{4,2}-\sqrt{2}\rho_{2,5})]\;,\\
    &\rho_{5,5}=\frac{\texttt{i}}{\gamma_{p}}[\sqrt{2}J(\rho_{5,4}-\rho_{4,5})+F(\rho_{5,2}-\rho_{2,5})]\;,\\
    &\rho_{4,4}=\frac{\texttt{i}}{2\gamma_{p}}(\sqrt{2}J(\rho_{4,5}-\rho_{5,4})+\sqrt{2}(\rho_{4,2}-\rho_{2,4})\;,\\
    &\rho_{4,6}=\frac{\texttt{i}}{2\gamma_{p}}(\sqrt{2}J(\rho_{4,5}-\rho_{5,6})+\sqrt{2}(\rho_{4,3}-\rho_{2,6})\;,\\
    &\rho_{6,4}=\frac{\texttt{i}}{2\gamma_{p}}(\sqrt{2}J(\rho_{6,5}-\rho_{5,4})+\sqrt{2}(\rho_{6,2}-\rho_{3,4})\;,\\
    &\rho_{6,6}=\frac{\texttt{i}}{2\gamma_{p}}(\sqrt{2}J(\rho_{6,5}-\rho_{5,6})+\sqrt{2}(\rho_{6,3}-\rho_{3,6})\; .
\end{eqnarray}
%\end{widetext}
from these closed set of equations we get a solution for
$\rho_{4,4}$ as
%\begin{widetext}
\begin{eqnarray}     \label{rho442cav}
    &\rho_{4,4}=\rho_{6,6}=\rho_{4,6}=\rho_{6,4}=   \nonumber \\
    &\frac{F}{2\gamma_{p}[(2\gamma_{p})^{2}+(2U)^{2}+(4J)^{2}]}  % \nonumber \\
    \times[\Im(\rho_{2,4})8\sqrt{2}(\gamma_{p}^{2}+U^{2}+2J^{2}+J U)+ %\nonumber \\
                -\Re(\rho_{2,4})8\sqrt{2}\gamma_{p} J+ \nonumber \\
                &+\Im(\rho_{2,5})8J(2J+U) + \Re(\rho_{2,5})8\gamma_{p} J]\;,
\end{eqnarray}
%\end{widetext}
where $\Re$ and $\Im$ indicate real and imaginary parts of the
respective elements. The explicit analytic expression for
$\rho_{4,4}$ can be calculated from those of the elements
$\rho_{2,2}$, $\rho_{1,4}$, $\rho_{1,5}$, $\rho_{2,4}$,
$\rho_{2,5}$, which have been obtained before. We notice that the
recursive procedure described can be properly generalized to
calculate the relevant density matrix elements also for generic
multi-site CQED systems.

\section{Numerical solution}\label{app:numerical}

{To solve Eqs.~(\ref{master:eq})-(\ref{L2C}) we use the finite-size
Fock-state matrix representation of all the operators. For any given
set of model parameters, the steady state density matrix can be
obtained by numerically searching for the eigenvector
$|\rho\rangle\rangle_{ss}$ corresponding to the eigenvalue
$\lambda_{ss}=0$ of the linear operator equation
\begin{equation}
\hat{L}|\rho\rangle\rangle=\lambda|\rho\rangle\rangle \, .
\end{equation}
In the latter, $|\rho\rangle\rangle$ is the density operator mapped
into vectorial form, and $\hat{L}$ is the linear matrix
corresponding to the Liouvillian superoperator on the right-hand
side of the master equation. If it exists, as it is always the case
for the parameters considered, the steady state solution is
unique.\cite{stenholm03pra} After recasting the vector
$|\rho\rangle\rangle_{ss}$ in matrix form, the relevant observable
quantities can be calculated as  $\langle O \rangle_{ss} = Tr\{
\hat{O}\rho_{ss}\}$. In this work, we kept  up to 6 photons per
cavity in the basis, which is sufficient for convergence due to the
weak driving conditions. Steady state results obtained in this way
have been successfully compared to the ones obtained from a full
time evolution of Eq.~(\ref{master:eq}) as a further check.}

\end{widetext}


\begin{thebibliography}{99}
\bibitem{gammap}
{ The polariton dissipation rate can be expressed through Eq.~(3) as
the weighted sum of the excitonic dissipation ($\gamma_x$) and bare
cavity photon ($\kappa$) dissipation rates, $\gamma_p=u^2 \gamma_x +
v^2 \kappa$.\cite{ciuti03sst} }
\end{thebibliography}
\end{document}